\documentclass[twocolumn,aps,prl,floatfix,longbibliography]{revtex4-2}

\usepackage{amsmath}
\usepackage{txfonts}
\usepackage{microtype}
\usepackage{graphicx}
\usepackage{color}
\usepackage{ulem}
\usepackage{bm}

\begin{document}

\title{ Atomic polarization and Stark-shift  in relativistic strong field ionization}

\author{Michael Klaiber}\email{klaiber@mpi-hd.mpg.de}
\affiliation{Max-Planck-Institut f\"ur Kernphysik, Saupfercheckweg 1, 69117 Heidelberg, Germany}
\author{Karen Z. Hatsagortsyan}\email{k.hatsagortsyan@mpi-hd.mpg.de}
\affiliation{Max-Planck-Institut f\"ur Kernphysik, Saupfercheckweg 1, 69117 Heidelberg, Germany}
\author{Christoph H. Keitel}
\affiliation{Max-Planck-Institut f\"ur Kernphysik, Saupfercheckweg 1, 69117 Heidelberg, Germany}

\date{\today}

\begin{abstract}

A relativistic analytical theory of strong field ionization applicable across the regimes of the deep-tunneling up to the over-barrier ionization (OTBI) is developed, incorporating the effects of the polarization of the atomic bound state and the Stark-shift in an ultrastrong laser field. The theory,  in particular,  addresses the order of magnitude discrepancy of the ionization yield in the relativistic regime calculated via the numerical solution of the Klein-Gordon equation [B. Hafizi \textit{et al}., Phys. Rev. Lett. 118, 133201 (2017)] with respect to the state-of-the-art quasiclassical theory of Perelomov-Popov-Terent'ev (PPT) for strong field ionization. The developed theory employs a Keldysh-like approach describing the ionization as an adiabatic quantum jump from the bound state to the continuum at a specific transition time, where the improved performance is achieved by accounting for the bound state distortion in the laser field. In the nonrelativistic limit, our theory reproduces the well-known fit to the numerical calculations for the OTBI rate via the Tong-Lin factor.  Realistic conditions for an experimental confirmation of the prediction of the present relativistic model versus  PPT-theory are also presented.

\end{abstract}

\maketitle

The experimental investigation of relativistic strong field ionization has been initiated since the pioneering experiment of Moore \textit{et al.} \cite{Moore_1999} more than 20 years ago at a laser intensity of $3\times 10^{18}$~W/cm$^2$, demonstrating the ionized electron ponderomotive acceleration. Later,  in more detailed atomic physics experiments, the observation of signatures of the atomic bound dynamics in  photoelectron momentum distributions (PMD) has been investigated
in relativistic laser fields \cite{Chowdhury_2001,Dammasch_2001,Yamakawa_2003,Maltsev_2003,Gubbini_2005,DiChiara_2008,
Palaniyappan_2008,Ekanayake_2013}. Presently ultrastrong laser fields up to the intensity of $10^{23}$~W/cm$^2$ are achievable \cite{Yoon_2021}, which provides a good perspective for extending the ionization explorations in the  relativistic regime.

The analytical treatment of strong field ionization, gained an impetus from the seminal Keldysh approach \cite{Keldysh_1965}, and developed into the quasiclassical Perelomov-Popov-Terent'ev (PPT) theory \cite{Smirnov_1966,Perelomov_1966a,Popov_1967,Perelomov_1967a,ADK,Popov_2004u}, as well as  into the strong field approximation (SFA) \cite{Keldysh_1965,Faisal_1973,Reiss_1980}, and have been generalized into the relativistic regime in \cite{Popov_1997,Mur_1998,Milosevic_2002r1,Milosevic_2002r2}, and  \cite{Reiss_1990,Reiss_1990b},  respectively. The PPT theory uses the quasiclassical wave function for the description of the tunneling part of the electron wavepacket
which is matched to the undisturbed exact bound state. The deficiency of the PPT theory is that the distortion of the bound state in the laser field during the ionization process is not taken into account. This is not essential during  tunneling ionization (TI)  and the PPT theory well approximates the experimental ionization yield, and has been applied for the calibration of ultrahigh laser intensities \cite{Hetzheim_2009,Bauke_2011,Ciappina_2019,Ciappina_2020}. However,  it becomes crucial in the over-the-barrier ionization (OTBI) regime, and PPT is known to significantly overestimate the ionization yield in the latter case \cite{Scrinzi_1999,Bauer_1999,Tong_2005,Lotstedt_2020}.
For the nonrelativistic regime, an adiabatic theory of  tunneling ionization has been developed accounting for the Stark-shift of the bound state \cite{Batishchev_2010,Tolstikhin_2010,Tolstikhin_2012,Trinh_2013}, applying perturbation theory  for the atomic Siegert states in a constant electric field~\cite{Siegert_1939}. The feasibility of the observation of relativistic features of the ionization yield is recently discussed in Refs.~\cite{Popruzhenko_2023a,Popruzhenko_2023c} (for the features of relativistic recollisions see Ref.~\cite{Kohler_adv} and Refs. therein).

Numerical investigation of the relativistic ionization dynamics via the Dirac equation,  with highly-charged ions (HCIs) and ultrastrong laser  fields has been carried out in Refs.~\cite{Hu_1999,Casu_2000,Hu_2002,Walser_2002,Mocken_2004b,Mocken_2008,Hetzheim_2009,Bauke_2011,Selsto_2009,
Vanne_2012,Fillion-Gourdeau_2012,Kjellsson_2017a,Telnov_2020,Hafizi_2017}. In particular, the total ionization yield via a 3D solution of the Klein-Gordon equation for hydrogen-like HCI has been calculated \cite{Hafizi_2017}. The surprising result is that the yield underestimates by an order of magnitude the prediction of the relativistic PPT theory, with the discrepancy increasing in the deep relativistic regime.  A hint on the possible reason for the observed discrepancy can be deduced from Ref.~\cite{Jones_2023}, where the role of the atomic state polarization and the Stark-shift of the bound state  in the nonrelativistic regime has been investigated for the tunneling ionization yield.
Do these effects compensate each other in the tunneling ionization  regime (but not for OTBI), explaining the good performance of PPT theory for tunneling?

In this letter, a  theory of strong field ionization in the relativistic regime is put forward, which incorporates the effects of the polarization of the atomic bound state and the Stark-shift of the bound state energy in an ultrastrong laser field.  We use a Keldysh-like approach describing the ionization as an adiabatic quantum jump from the bound state to the continuum at a specific transition time and calculate the ionization probabilities based on the overlap integral of the continuum and the bound state at this time. In contrast to the common PPT theory, we account for the bound state distortion in the laser field using generalized eikonal (GEA) theory \cite{Kaminski_1984,Avetissian_1999,Kaminski_2015}. The latter improvement is crucial to reproduce the ionization rates from the deep-tunneling up to the OTBI regime, in particular, to explain the order of magnitude discrepancies of the numerical result of Ref.~\cite{Hafizi_2017}  for the relativistic ionization with respect to PPT theory. In light of the present theory,  the recent experimental results of Refs.~\cite{Yandow_2024} are analyzed and the deviation from PPT is explained. We discuss the conditions for an experimental confirmation of the predictions of the present  relativistic model versus  PPT-theory. In the nonrelativistic regime the theory allows us to derive analytically the fitting formula of Ref.~\cite{Tong_2005}   (Tong-Lin factor) to the OTBI yield via numerical solutions of the time-dependent Schr\"odinger equation (TDSE).

Our approach for calculating  strong field ionization probabilities is based on the use  of the modified continuum and the bound states. The wave function of the electron is given by the exact time-evolution operator (TEO) $U(t,t')$ of the system:
$|\psi(t)\rangle=U(t,t^a)|\psi(t^a)\rangle$, with the initial condition $|\psi(t^a)\rangle=|\psi^a_0(t^a)\rangle$, where $\psi^a_0$ is the unperturbed atomic bound state at the turn on  $t^a$ of the laser field. The ionization amplitude $m_\mathbf{p}$ is derived by a projection of $\psi$ on the exact continuum state $\psi^f_\mathbf{p}$ with asymptotic momentum $\mathbf{p}$ at the asymptotic time~$t^f$:
\begin{eqnarray}
m_\mathbf{p}=\langle \psi^f_\mathbf{p}(t^f)|U(t^f,t^a)|\psi(t^a)\rangle,
\end{eqnarray}
which determines the differential ionization probability: $ dw/d^3\mathbf{p} =|m_\mathbf{p}|^2$.
We approximate the exact TEO  following the Keldysh-approach \cite{Keldysh_1965},  assuming that in the beginning of the ionization process the atomic potential dominates the dynamics, whereas in the end -- the laser field:
\begin{eqnarray}
U(t^f,t^a)=U^f(t^f,t^*)U^a(t^*,t^a),
\end{eqnarray}
where $U^f$ is the TEO  with a perturbative treatment of the atomic potential $V$, and $U^a$ is the TEO with a perturbative treatment of the laser field.
The time $t^*$ is the instant when the transition between the two approximations of the exact TEO takes place. In accordance with the adiabatic transitions theory \cite{Landau_3,Dykhne_1962,Briggs_2000,Dimitrovski_2003}, the transition time is  determined by the condition of the quasienergy equality  of the two adiabatically evolved states. The quasienergy of a state is defined as $\varepsilon=-\partial_t{S}$, with the action $S$: $\psi=\exp(iS)$. Thus, the transition time from the atomic state modified in the laser field [$\psi^a=\exp(iS^a)$]  to the continuum state in the laser field, modified by the Coulomb potential of the atomic core [$\psi^f=\exp(iS^f)$], is found via
\begin{eqnarray}
\partial_t {S}^{a}(t^*)=\partial_t{S}^{f*}(t^*).
\label{t*}
\end{eqnarray}
The adiabatic approximation is valid when the typical timescale of the perturbation exceeds that of the state. For the strong field ionization this implies that the laser period exceeds the atomic evolution time: $\omega \ll I_p$, with the laser frequency $\omega$, and the atomic ionization potential $I_p$. The ionization amplitude obtains then a simple form of an overlap integral:
\begin{eqnarray}
\label{m_p}
m_\mathbf{p}=\langle \psi^{f}_{\mathbf{p}}(t^*)|\psi^{a}(t^*)\rangle.
\end{eqnarray}
To derive the modified continuum and bound states wavefunctions, we employ  GEA describing the eikonal $S$ perturbatively. For the continuum wavefunction the atomic potential is a perturbation in describing the phase, and for the bound state -- the interaction with the laser field.

 \textit{Nonrelativistic regime.} Let us begin with the nonrelativistic case. The Schr\"odinger equation for the eikonal $S$ reads:
\begin{eqnarray}
\label{eikonal}
-\partial_t S= \left({\boldsymbol \nabla} S \right)^2/2-i \Delta S/2+V(r)+\mathbf{r}\cdot \mathbf{E}(t).
\end{eqnarray}
The description of the continuum motion is modified, treating the atomic potential $V(r)=-Z/r$, with the charge of the atomic core $Z$, by the perturbation theory in Eq.~(\ref{eikonal}): $S^f=S^f_0+S_1^f$, where the unperturbed eikonal [at $V=0$] corresponds to the nonrelativistic Volkov wavefunction $\psi^{f}_{\mathbf{p},0}$~\cite{Becker_2002}: $S^f_0= [\mathbf{p}+\mathbf{A}(t)]\cdot \mathbf{r}+\int_tdt' [\mathbf{p}+\mathbf{A}(t)]^2/2$, and the perturbed eikonal  $S_1^f$ fulfills the equation
\begin{eqnarray}
\label{eikonal_1}
-\partial_t S_1^f-[\mathbf{p}+\mathbf{A}(t)]\cdot  {\boldsymbol \nabla} S_1^f +i \Delta S_1^f/2=V(r).
\end{eqnarray}
The solution of Eq.~(\ref{eikonal_1}) is known from the GEA theory \cite{Kaminski_2015}:
\begin{eqnarray}
\label{S1}
S_1^{f}(\mathbf{r},t)=-Z\int_t ds \frac{\text{erf}\left[\sqrt{\frac{(\mathbf{r}+\mathbf{p}(s-t)+\boldsymbol{\alpha}(s)-\boldsymbol{\alpha}(t))^2}{-2i(s-t)}}\right]}{\sqrt{(\mathbf{r}+\mathbf{p}(s-t)+\boldsymbol{\alpha}(s)-\boldsymbol{\alpha}(t))^2 }},
\end{eqnarray}
where $\boldsymbol{\alpha}(t)=\int dt\mathbf{A}(t)$ is the displacement in the laser field, and  $S_1^f$   describes the modification of the tunneling barrier due to the Coulomb field,  enhancing the ionization probability~\cite{Perelomov_1967a}.

\begin{figure}[b]
  \begin{center}
 \includegraphics[width=0.4\textwidth]{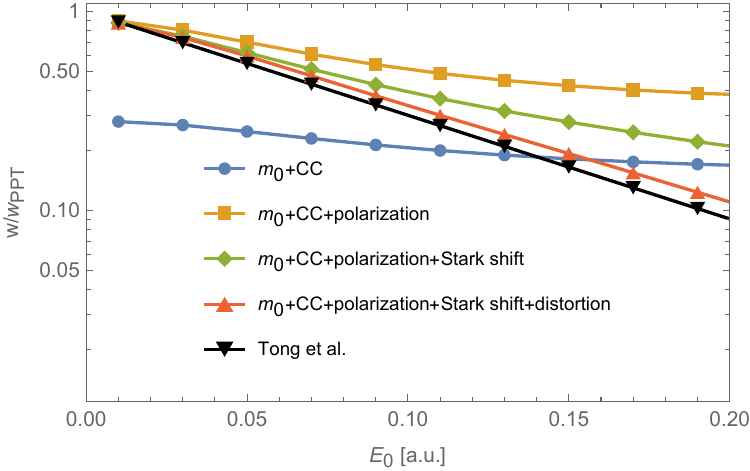}
 \caption{ The ratio of the ionization rate $w$ to that of PPT-theory $w_{\rm PPT}$ for hydrogen ($\kappa=1$): (blue circles) our model with only Coulomb corrections (CC) via $S_1^f$; (orange squares) our model with CC and the atomic polarization (via $S_1^f$, $S_1^a$ and $S_2^a+\varepsilon^s t$); (green diamonds) our model with CC, the atomic polarization, and the Stark-shift; (red triangles) our model with all corrections including the bound state distortion; (black inverted triangles) the  Tong-Lin-factor \cite{Tong_2005}.}
  \label{Tong}
\end{center}
\end{figure}

In a similar manner, the description of the bound state dynamics is modified, treating the interaction with the laser field  as a perturbation in the eikonal in Eq.~(\ref{eikonal}): $S^a=S^a_0+S_1^a$,
 where the unperturbed eikonal  corresponds to the free atomic wavefunction $S^{a}_0(\mathbf{r},t)=i\kappa r-i\left(Z/\kappa-1\right)\ln(\kappa r)-i\ln(\sqrt{\kappa^3/\pi})+I_pt+c^a_0$ \cite{Bethe_1957}, with the atomic velocity $\kappa=\sqrt{2I_p}$, and the normalization constant $c^a_0$.
  The perturbed eikonal  $S_1^a$  describes the polarization of the atomic state in the laser field and fulfills
\begin{eqnarray}
\label{eikonal_a}
-\partial_t S_1^a-{\boldsymbol \nabla} S_0^a \cdot  {\boldsymbol \nabla} S_1^a +i \Delta S_1^a/2= \mathbf{r}\cdot \mathbf{E}(t).
\end{eqnarray}

First, we consider the quasistatic case $\mathbf{E}(t)=-\hat{\mathbf{x}}E_0=\rm const$ (with $p_x=0$ without loss of generality).
Introducing parabolic coordinates  $u=\sqrt{r+x}$ and $v=\sqrt{r-x}$, the  solution close to the origin is, see the Supplemental Materials (SM) \cite{SM}:
\begin{eqnarray}
\label{S1a}
 S^a_1=-\frac{i E_0  }{8 \kappa^2}\left[u^2 (4 +  u^2\kappa)-v^2 (4 + v^2\kappa)\right].
\end{eqnarray}
We calculate also the second  order correction, see SM \cite{SM}, which is important to describe the Stark-shift: $S^a_2=-\frac{i E_0^2  }{96 \kappa^4}  \left[ u^4  (21 + 2 u^2\kappa) +v^4   (21 + 2  v^2\kappa)\right] - \varepsilon^s t$,
with the Stark-shift energy $\varepsilon^s=-9E_0^2/4\kappa^4$. Thus, the modified bound wave function including the atomic polarization and the Stark-shift is
$\psi^a \approx c^a_2\psi^ a_0[1+i S_1^a-(S_1^{a})^2/2+iS_2^a]$,
with the normalization constant $c^a_2\approx1-31E_0^2/(4\kappa^6)$,  where we expanded the exponent as $S_{1,2}^a\ll 1$ at $r\sim 1/\kappa$. With the replacement $E_0\rightarrow E(t)$, the solutions above for  $S_{1,2}^a$  are valid also in time-dependent fields at $ (E_0/E_a)\gamma^2/\sqrt{1+\gamma^2}\ll 1$, with the atomic field $E_a\equiv\kappa^3$, which is equivalent to $\omega/I_p\ll 1$ at a large Keldysh-parameter $\gamma=\kappa\omega/E_0$ \cite{Keldysh_1965}.

The ionization amplitude is calculated via Eq.~(\ref{m_p}) numerically with the modified wavefunctions $\psi^f$ and $\psi^a$. The integration over the coordinate in Eq.~(\ref{m_p}) is extended up to the tunnel exit $x_e=I_p/E_0$, as the tail of the wavefunction out of the barrier cannot contribute to ionization, see  SM~\cite{SM}.
The calculated ionization rate ($w$), highlighting different contributions,  is illustrated in Fig.~\ref{Tong}.  It is remarkable that it reproduces accurately the Tong-Lin fitting factor \cite{Tong_2005} of the ionization yield via numerical TDSE solutions with respect to PPT: $(w/w_{\rm PPT})_{\rm nr}=\exp[-12(Z^2/\kappa^2)(E_0/E_a)]$, with the PPT rate $w_{\rm PPT}$ \cite{Popov_2004u}.
We estimate analytically the Tong-Lin factor with our model at $f\equiv E_0/E_a\ll 1$, see SM \cite{SM}:
\begin{eqnarray}
\label{m1xe}
\left(\frac{w}{w_{\rm PPT}}\right)_{\rm nr}&=&\frac{15}{8}+\frac{4}{3} \sqrt{\frac{2f}{\pi }} -\frac{e^{ -\frac{1}{8f} }}{8\sqrt{2 \pi  }f^{3/2} } \left(\frac{1}{4}+\frac{29f}{3} +\frac{271 f^2}{3  } \right).\nonumber\\
 \label{m11}
\end{eqnarray}
We highlight the following contributions in Fig.~\ref{Tong}: the Stark-shift described via the eikonal term $\varepsilon^s t$, the polarization effect as a shift of the bound state toward the tunnel exit (via  $S_1^a$ and $S_2^a+\varepsilon^s t$), and the the polarization effect of the bound state distortion (via the factor $c^a_2$). The shift of the bound state increases the  ionization rate (by a factor of $\sim 4$) up to  the PPT value for weak fields. This is because the PPT rate implicitly includes this polarization effect via the field dependence of the matching coordinate $x_s=\sqrt{\kappa/E_0}$ of the undisturbed bound wavefunction to the continuum. This is the reason of the good performance of the PPT theory in the tunneling regime. In strong fields, the bound state distortion  and the Stark-shift  decrease the rate away from the PPT  result with the respective scaling: $\sim -15E_0^2/E_a^2$, and $\sim-5 E_0/E_a$, according to our model.

 \textit{Relativistic regime. Klein-Gordon equation. } Firstly, we look for the solution of Klein-Gordon equation using the ansatz $\psi=\exp(i{\cal S})$, where the eikonal ${\cal S}$ fulfills the equation  \cite{Avetissian_1999}:
\begin{eqnarray}
\label{Dirac}
 -i\partial^2 {\cal S}(x)+[\partial {\cal S}(x)+A(\eta)/c+V(x)/c]^2=c^2,
\end{eqnarray}
with the four-coordinate $x_\mu=(ct, \mathbf{r})$, $\partial\equiv \partial/\partial x_\mu$, the laser four-vector potential $A(\eta)=(\mathbf{r}\cdot \mathbf{E}(\eta), -\hat{\mathbf{k}} (\mathbf{r}\cdot \mathbf{E}(\eta))$ in the G\"oppert-Mayer gauge, $\hat{\mathbf{k}}=\mathbf{k}/|\mathbf{k}|$, the laser wavevector $\mathbf{k}$, and the atomic four-potential $V(x)=(V( r),0,0,0)$. The  eikonal is derived by perturbation theory ${\cal S}={\cal S}_0+{\cal S}_1$, either with respect to $V$ for ${\cal S}^f$, or with respect to $A(\eta)$ for ${\cal S}^a$.
The  unperturbed eikonal of the relativistic continuum state ${\cal S}_0^f$ is represented by the phase of the relativistic Volkov wavefunction \cite{Volkov_1935}: ${\cal S}_0^f (x,y,z,,t)= [p_x+A(\eta)] x+p_y y-c\Lambda z  +\int_\eta ds\tilde{\varepsilon}(s)$.
Here, the laser wave propagates along the $z$-axis, and is linearly polarized along the $x$-axis, $\eta= t-z/c$, $\Lambda\equiv  \varepsilon/c^2-p_z/c$ is the motion integral in a plane wave field, $\varepsilon=\sqrt{c^4+c^2p^2}$ is the electron energy, and $\tilde{\varepsilon}(s)=\varepsilon+[p_xA(\eta)+A(\eta)^2/2]/\Lambda$ is the  energy in the laser field.
The GEA correction to the eikonal of the continuum state ${\cal S}_1^f$ fulfills the equation
\begin{eqnarray}
\label{Dirac1f}
 i\partial^2 {\cal S}_1^f(x)+2[\partial {\cal S}_0^f(x)+A(\eta)/c][\partial {\cal S}_1^f(x)+V(x)/c]=0.
\end{eqnarray}
After coordinate transformation to $(\eta,\mathbf{r})$, Eq.~(\ref{Dirac1f})  is solved via  Fourier transformation, see SM \cite{SM}. In the case of a quasistatic field, using the zeroth-order solution  in velocity gauge
${\cal S}^f_0=-c\Lambda z-\varepsilon \eta-E_0^2\eta^3/(6\Lambda)$,
and assuming  $q_z\ll c$, the solution for the Fourier component is derived \cite{SM}, leading to
\begin{eqnarray}
{\cal S}_1^{f}(\eta,\mathbf{r})\approx\int_\eta ds \frac{-Z \tilde{\varepsilon}(s)\text{erf}\left[\sqrt{\Lambda\frac{r(s,\eta)^2}{-2i(s-\eta)}}\right]}{c^2 \Lambda r(s,\eta)},
\end{eqnarray}
with the relativistic trajectory $r(s,t)=\{ [x+ (\alpha(s)-\alpha(\eta))/\Lambda ]^2+ [y+ p_y(s-\eta)/\Lambda ]^2 + [z+(p_z(s-\eta)/\Lambda + (\beta(s)-\beta(\eta))/(c\Lambda^2) ]^2 \}^{1/2}$, and $\beta=\int d\eta A^2(\eta)/2$.

\begin{figure}[b]
  \begin{center}
\includegraphics[width=0.4\textwidth]{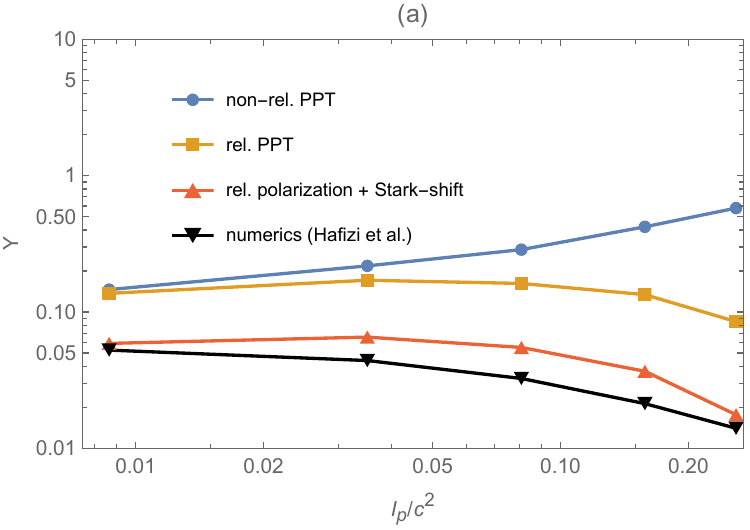}
\includegraphics[width=0.4\textwidth]{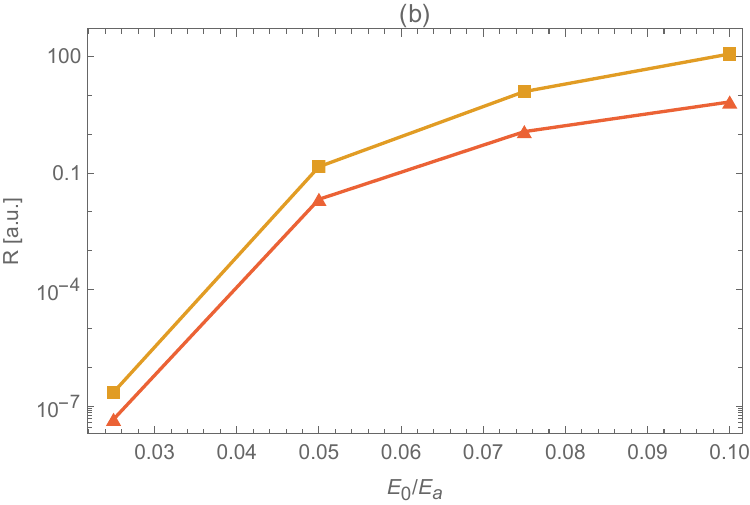}
 \caption{(a) Ionization yield $Y$ via the Klein-Gordon equation vs $I_p/c^2$, (b) ionization rate $R$ via the Dirac equation vs $E_0/E_a$ for $I_p/c^2=0.2$.; The numerical calculation  of  Ref.~\cite{Hafizi_2017} (black inverted triangles), relativistic PPT \cite{Milosevic_2002r1} (orange squares), non-relativistic PPT (blue circles),  and our matching method with relativistic polarization and Stark-shift (red triangles);  The curves  in Panel (a) for the ionization energies considered herein $I_p/c^2 = 0.00866, 0.0351, 0.0809, 0.158, 0.259$; the corresponding  peak electric fields are $E_0/E_a= 0.0630, 0.0635, 0.0643, 0.0662, 0.0684$, respectively~\cite{Hafizi_2017}. Note that the normalized electric field is nearly constant across the range of ionization energies.}
\label{rel}
\end{center}
\end{figure}

The unperturbed eikonal for atomic Klein-Gordon wavefunction is ${\cal S}^a_0=S^a_0-i[(1-I_p/c^2)^2-1]\ln (\kappa r) -i \ln {\cal C}^a_0$, according to the relativistic bound state  \cite{Bethe_1957}, where $S^a_0$ is the nonrelativistic term, and  ${\cal C}^a_0=\sqrt{1/2c}2^{-1 + \epsilon_0^2}[\epsilon_0 \Gamma(2 \epsilon_0^2)]^{-1/2}$, with $\epsilon_0\equiv 1-I_p/c^2$.
The correction to the bound state eikonal fulfills the equation:
\begin{eqnarray}
\label{Dirac1a}
-i\partial^2 {\cal S}_1^a(x)+2[\partial {\cal S}_0^a(x)+V(x)/c][\partial {\cal S}_1^f(x)+ A(\eta)/c]=0,
\end{eqnarray}
and similar equation for ${\cal S}_2^a$.  The equations for ${\cal S}_{1,2}^a$ are solved
in a quasiclassical expansion~\cite{Klaiber_2013_II}, up  to the next to leading order in $\hbar$ \cite{SM}. The main corrections to the bound state come from the atomic polarization and Stark-shift due to the laser electric field. Additionally  new terms in the relativistic regime occur due to the laser magnetic field, and the electron mass correction:
\begin{eqnarray}
{\cal S}^a_1&=&S^a_1+ix  \left\{   -E_a (1 - I_p/c^2) r/(2\kappa) +i E_a z  /(2c) +
   r \kappa^2/2  \right.\nonumber\\
   &+&\left.  \kappa  - (E_a /\kappa^2) [(1 - I_p/c^2)^3 + Z \kappa/c^2]  \right\} (E_0/E_a),
\end{eqnarray}
where $E_a=\sqrt{3} \sigma^3/(1 + \sigma^2) c^3$ is the relativistic atomic field~\cite{Milosevic_2002r2}, with $\sigma=\sqrt{1 + \epsilon_0^2/2 - \epsilon_0 \sqrt{4 + \epsilon_0^2/4}}$. The  correction ${\cal S}^a_2$ is calculated in a similar way, see SM \cite{SM}.

 \textit{Dirac equation}. The wavefunction for Dirac equation is looked for via the ansatz $\psi=u\exp(i{\cal S})$, with the spinorial part $u$, and the eikonal ${\cal S}$ of the Klein-Gordon equation.
We assume that for the continuum state $\psi^f=u_0^f\exp(i{\cal S}^f)$,  with the spinor $u_0^f$ of the Volkov wavefunction~\cite{Volkov_1935}. In this ansatz, we neglect the sub-barrier spin effects due to the atomic potential during tunneling, which are of the order of $ (\kappa/c)^3$  \cite{Klaiber_2013_II}.

The unperturbed relativistic bound state wavefunction for the Dirac equation is given by \cite{Bethe_1957}
${\cal S}^a_0=S^a_0-i(I_p/c^2)\ln\left(\kappa r\right)-i \ln({\cal C}^a_0)$, with ${\cal C}^a_0=2^{-1 + \epsilon_0} \sqrt{(1 + \epsilon_0)/\Gamma(1 + 2 \varepsilon_0)}$.
 We choose the Dirac eikonal for the atomic wavefunction to be identical to the Klein-Gordon one ${\cal S}^a$ and find the spinorial corrections. The first order spinorial correction in $\hbar$ and $E_0/E_a$  reads:

\begin{gather}
 u_1^a=  i\frac{E_0 t}{2c}u^a_0+\frac{E_0}{E_a}
  \begin{bmatrix}
  \frac{ I_p^{3/2} (-iz+3x)}{\sqrt{2}c^2}  \\
  \frac{i I_p^{3/2} y}{\sqrt{2}c^2}-\frac{I_pr}{c} \\
 -\frac{i I_p x y}{2 rc}-\frac{I_pr}{c}+i\frac{ I_p^{3/2} y}{\sqrt{2}c^2}  \\
 -\frac{-I_p (z^2 + i z x + y^2)}{2 rc}
   \end{bmatrix},
\end{gather}
where we applied $1/c$-expansion up to second order. Similarly, we find $u_2^a$ \cite{SM}. Using the modified atomic and continuum wavefunctions in the relativistic regime,
$\psi^a={\cal C}^a_2[\psi^a_0(1+i {\cal S}_1^a-({\cal S}_1^{a})^2/2+{\cal S}_2^{a})+\exp(i{\cal S}^a_0)(u^a_1+i{\cal S}^a_1u^a_1+u^a_2)]$,
the ionization amplitude of Eq.~(\ref{m_p}) for Klein-Gordon and Dirac equations is calculated  numerically at the most probable momentum $\mathbf{p}=(0,0,I_p/3c)$ \cite{Klaiber_2013}, see  Fig.~\ref{rel}, where the ionization yield in different approximations are compared. We also estimated analytically the relativistic analogue of the Tong-Lin-factor (the correction factor to the PPT ionization rate), see SM \cite{SM}:
\begin{eqnarray}
\left(\frac{w}{w_{\rm PPT}}\right)_{\rm rel}  \approx \left(\frac{w}{w_{\rm PPT}}\right)_{\rm nr}e^{ - \frac{2\kappa^2}{c^2} }\approx \exp\left(-\frac{12Z^2}{\kappa^2}\frac{E_0}{E_a}-\frac{2\kappa^2}{c^2}\right).
\end{eqnarray}

 \begin{figure}[b]
  \begin{center}
\includegraphics[width=0.4\textwidth]{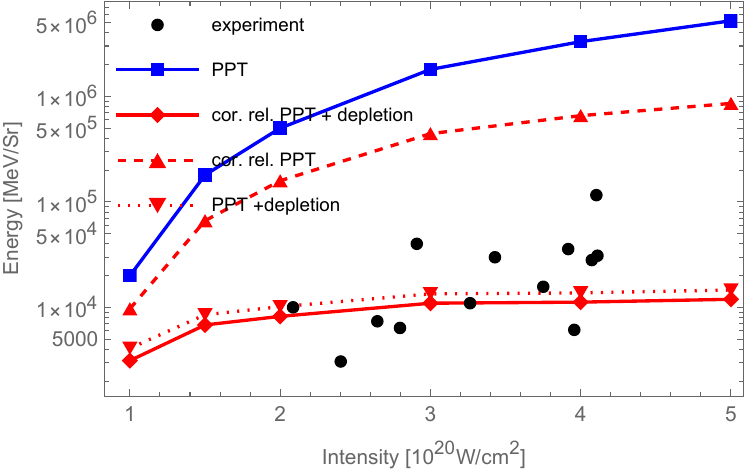}
\includegraphics[width=0.4\textwidth]{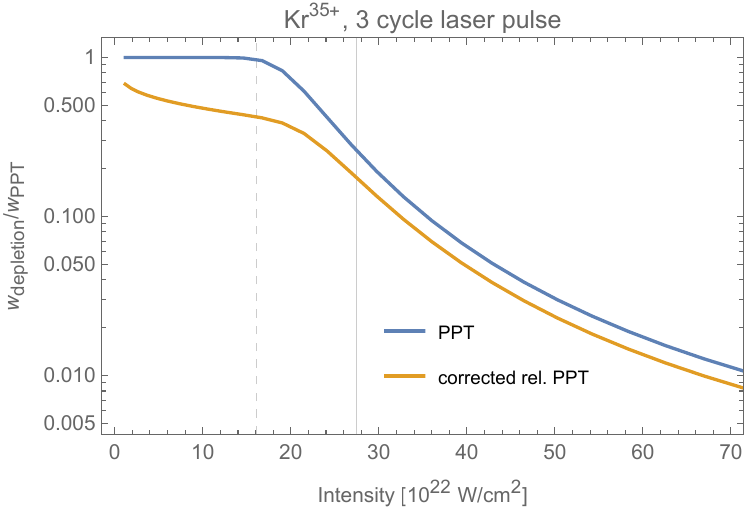}
 \caption{(a) Integrated electron energy vs laser intensity: the experimental results of Refs~\cite{Yandow_2024} with the 1 mm shield (black cycles). The PPT-theory from Refs~\cite{Yandow_2024} (blue-solid, squares), PPT with depletion (red-dotted, inverted triangles), the corrected PPT-theory with the present model (red-dashed, triangles), our model including the depletion effect of the ground state (red, diamonds).
 (b) The  ratio of the ionization yield for Kr$^{35+}$ with respect to PPT  in a 3 cycle laser pulse for (blue) the PPT yield with depletion, (red) our model with depletion. The grid lines show the OTBI threshold (solid), and the saturation intensity for Kr$^{34+}$ (dashed). }
\label{ditmire}
\end{center}
\end{figure}

According to Fig.~\ref{rel}, our theory provides a good agreement for the yield with the results of the numerical solution of the Klein-Gordon equation of Ref.~\cite{Hafizi_2017}, while the standard relativistic PPT theory overestimates it by more than an order of magnitude.  The role of the different polarization effects (bound state shift and distortion) and the Stark-shift is similar to the nonrelativistic case [Fig.~\ref{Tong}]. The main characteristic feature of the relativistic yield is the decreasing of the yield at large $I_p/c^2$. This stems from the relativistic mass shift effect, which decreases the size and the polarization of the atomic bound state at large $I_p/c^2$.  The latter contributes to the deviation of the result with our model with respect to the relativistic PPT.

 \textit{Analysis of the experiment of Ref.~\cite{Yandow_2024}.} Recently an experiment on tunneling ionization from the $K$-shell of neon in the relativistic regime has been carried out with a laser intensity exceeding $10^{20}$ W/cm$^2$ \cite{Yandow_2024}. The integrated electron energy is measured. The authors provide also Monte Carlo simulations employing the  PPT rates and conclude that the PPT theory (as well as the so-called barrier suppression ionization model \cite{Augst_1991}) overestimates the ionization yield. However, it appears that the depletion of the atomic state is not taken into account in these simulations. We give an estimation of the experimental results using the theoretical method of this paper and include also the depletion effect. Our qualitative estimation consists in multiplying the PPT-curve in Fig.~\ref{ditmire}(a) by the correction factor according to our theory $(w/w_{\rm PPT})_{\rm rel}$. The depletion is accounted for by evaluating the correction factor for the laser field corresponding to the ionization saturation time $t_d$, determined from the ionization yield $Y(t)$ via $\ddot{Y}(t_d)=0$, see SM \cite{SM}. When the depletion of the bound state is included,  both the PPT theory and our method fit the experimental result within the experimental error [Fig.~\ref{ditmire}(a), the red solid  and the red dotted lines], Thus, the experimental result of Refs.~\cite{Yandow_2024} cannot distinguish between PPT and our method. This is because of the domination of the depletion  in  a long laser pulse duration in the experiment (25 cycles). With a shorter laser pulse, the deviation of our method from PPT will be measurable near the OTBI threshold, as the example of krypton HCI ionization in a 3-cycle ultraintense laser pulse is demonstrated in Fig.~\ref{ditmire}(b). We see that near the OTBI threshold, the accounting of the bound state distortion in the laser field can result in a decrease of the ionization yield by more than 2 times.

Concluding, we put forward a simple model for relativistic ionization with an important ingredient of accounting for the bound state polarization and the Stark-shift. With this  modifications in the adiabatic transition model, the deviation of the numerical and experimental results from the  PPT theory is explained. In the recent experiment of Ref.~\cite{Yandow_2024} the depletion of the bound state is the dominating factor to account for the deviation from PPT theory. The role of the depletion can be avoided using a shorter laser pulses, when the prediction of our model deviating from PPT can be confirmed.

We thank Andrew Yandow for providing the experimental data. MK acknowledges fruitful discussions with John~S.~Briggs.

\bibliography{strong_fields_bibliography}

 \end{document}